\documentclass[conference,10pt]{IEEEtran}
\usepackage{cite}
\ifCLASSINFOpdf
   \usepackage[pdftex]{graphicx}
\else
\fi
\usepackage[font={small}]{caption}
\usepackage{amsmath}
\usepackage{amssymb}
\usepackage{xcolor}
\usepackage[nolist,nohyperlinks]{acronym}
\usepackage{algorithm}
\usepackage[noend]{algpseudocode}
\usepackage{url}
\newcommand{\E}[1]{\mathbb{E}\left\{#1\right\}}

\acrodef{UE}{user equipment}
\acrodef{BS}{base station}
\acrodef{Tx}{transmitter}
\acrodef{Rx}{receiver}
\acrodef{CFO}{carrier frequency offset}
\acrodef{ULA}{uniform linear array}
\acrodef{OFDM}{orthogonal frequency-division multiplexing}
\acrodef{AWGN}{additive white Gaussian noise}
\acrodef{LO}{local oscillator}
\acrodef{MAP}{maximum \textit{a posteriori}}
\acrodef{mmWave}{millimeter wave}
\acrodef{subTHz}{sub-terahertz}
\acrodef{LOS}{line-of-sight}
\acrodef{NLOS}{non-line-of-sight}
\acrodef{MCMC}{Markov chain Monte Carlo}
\acrodef{VB}{variational Bayes}
\acrodef{EP}{expectation propagation}
\acrodef{KL}{Kullback-Leibler}
\acrodef{SBL}{sparse Bayesian learning}
\acrodef{RV}{random variable}
\acrodef{PDF}{probability density function}
\acrodef{EM}{expectation maximization}
\acrodef{5G}{fifth generation}
\acrodef{iid}{independent identically distributed}
\acrodef{ML}{maximum likelihood}
\acrodef{SCA}{successive convex approximation}
\acrodef{MM}{majorization-minimization}
\acrodef{ADMM}{alternating direction method of multipliers}
\acrodef{SNR}{signal-to-noise ratio}
\acrodef{AoA}{angle of arrival}
\acrodef{AoD}{angle of departure}
\acrodef{MSE}{mean squared error}
\acrodef{RMSE}{root mean squared error}
\acrodef{ReMSE}{relative mean squared error}
\acrodef{PSO}{particle swarm optimization}
\acrodef{NM}{Nelder-Mead}
\acrodef{B5G}{beyond 5G}
\acrodef{6G}{sixth generation}
\acrodef{MIMO}{multiple-input multiple-output}
\acrodef{LLN}{law of large numbers}
\acrodef{LGR}{large grid regime}
\acrodef{LPA}{long pilot approximation}
\acrodef{MC}{Monte Carlo}
\acrodef{ToF}{time of flight}
\acrodefplural{TOF}[TOFs]{times of flight}
\acrodef{CACC}{cross-antenna cross-correlation}
\acrodef{MMSE}{minimum mean square error}
\acrodef{DTFT}{discrete-time Fourier transform}
\acrodef{DFT}{discrete Fourier transform}
\acrodef{FFT}{Fast Fourier transform}
\acrodef{2P2T}{two precoders two transmissions}
\acrodef{TVP}{time-varying precoder}
\acrodef{RF}{radio frequency}
\acrodef{PSF}{point spread function}
\acrodef{DBSCAN}{density-based spatial clustering of applications with noise}
\acrodef{SSE}{Sum Square Error}
\acrodef{TDMA}{time division multiple access}
\acrodef{FDMA}{frequency division multiple access}
\acrodef{CDMA}{code division multiple access}
\acrodef{BER}{bit error rate}
\acrodef{MUSIC}{multiple signal classification}
\acrodef{URA}{uniform rectangular array}
\acrodef{JCS}{joint communication and sensing}
\acrodef{NPNT}{N-precoders N-transmissions}
\acrodef{SOR}{successive over-relaxation}
\acrodef{COMPAS}{concurrent mapping, positioning, and synchronization}
\acrodef{SLAM}{simultaneous localization and mapping}
\acrodef{AECD}{alternating exact coordinate descent}
\acrodef{ISAC}{integrated sensing and communications}
\acrodef{JSC}{joint sensing and communication}
\acrodef{FR2}{frequency range 2}
\acrodef{CSI}{channel state information}
\acrodef{SAGE}{space-alternating generalized expectation-maximization}
\acrodef{PCE}{parametric channel estimation}
\acrodef{PEB}{performance error bound}
\acrodef{DFRC}{dual-functional radar communications}
\acrodef{CRLB}{Cramér–Rao lower bound}
\acrodef{FIM}{Fisher information matrix}
\acrodef{OMP}{orthogonal matching pursuit}
\acrodef{AMP}{approximate message passing}
\acrodef{MAE}{mean absolute error}
\acrodef{OTFS}{orthogonal time frequency space}
\acrodef{I}{in-phase}
\acrodef{Q}{quadrature}
\acrodef{ADC}{analog-to-digital converter}
\acrodef{DAC}{digital-to-analog converter}
\acrodef{CDF}{cumulative distribution function}
\acrodef{CLT}{central limit theorem}
\acrodef{i.i.d.}{independent and identically distributed}
\acrodef{MSQE}{mean squared quantization error}
\acrodef{LM}{Lloyd-Max}
\acrodef{KS}{Kolmogorov–Smirnov}
\acrodef{XL-MIMO}{extremely large-scale MIMO}
\acrodef{PA}{power amplifier}
\acrodef{CS}{compressive sensing}
\acrodef{CP}{canonical polyadic}
\acrodef{HOSVD}{higher order singular value decomposition}
\acrodef{QPSK}{quadrature phase shift keying}
\acrodef{CPI}{coherent processing interval}
\acrodef{ACE}{adaptive coherence estimator}
\acrodef{CFAR}{constant false alarm rate}
\acrodef{ICM}{internal clutter motion}
\acrodef{LS}{least squares}
\acrodef{CPE}{common phase error}
\acrodef{SCNR}{signal-to-clutter-plus-noise ratio}
\acrodef{NMS}{non-maximum suppression}
\acrodef{TDD}{time-division duplex}
\acrodef{MoM}{method of moments}
\acrodef{CO}{coherent}
\acrodef{NC}{non-coherent}
\acrodef{DwC}{dwell-coherent}
\acrodef{FR1}{frequency~range~1}
\acrodef{GPU}{graphics processing unit}
\acrodef{PoC}{proof of concept}
\acrodef{GNSS}{global navigation satellite system}
\acrodef{UAV}{unmanned aerial vehicle}

\usepackage{subcaption}
\usepackage{diagbox}
\usepackage{multicol}
\usepackage{mathtools}
\usepackage{stfloats}

\makeatletter
\g@addto@macro\normalsize{%
  \setlength\abovedisplayskip{4.0pt plus 2pt minus 1pt}%
  \setlength\belowdisplayskip{4.0pt plus 2pt minus 1pt}%
  \setlength\abovedisplayshortskip{0pt plus 2pt}%
  \setlength\belowdisplayshortskip{2.0pt plus 1pt}}
\def\section{\@startsection{section}{1}{\z@}{1.25ex plus 1ex minus 0.4ex}%
{0.55ex plus 0.5ex minus 0ex}{\normalfont\normalsize\centering\scshape}}%
\def\subsection{\@startsection{subsection}{2}{\z@}{1.15ex plus 1ex minus 0.4ex}%
{0.50ex plus 0.5ex minus 0ex}{\normalfont\normalsize\itshape}}%
\makeatother

\newcommand{\chg}{\color{blue}} 
\renewcommand{\chg}{}

\begin{document}

\title{6G ISAC Drone Detection: Signal Processing and Experimental Validation at FR1}

\author{\IEEEauthorblockN{Enrique~T.~R.~Pinto\IEEEauthorrefmark{1}\IEEEauthorrefmark{2}, Marcus~Henninger\IEEEauthorrefmark{2}, Lucas~Giroto\IEEEauthorrefmark{2}, Markku~Juntti\IEEEauthorrefmark{1}, and Silvio~Mandelli\IEEEauthorrefmark{2}}
\IEEEauthorblockA{University of Oulu, Finland\IEEEauthorrefmark{1};         Nokia Bell Labs Stuttgart, Germany\IEEEauthorrefmark{2} \\
\{name.surname\}@oulu.fi\IEEEauthorrefmark{1}, \{name.surname\}@nokia-bell-labs.com\IEEEauthorrefmark{2}}
}

\markboth{}%
{}

\maketitle

\noindent
\begin{abstract}
Integrated sensing and communications (ISAC) enables mobile networks to detect passive targets such as drones at minimal additional cost. However, in ISAC, a small drone must be detected against strong ground clutter with the limited radio resources left by communications. In this paper, we propose two ISAC target detectors: a coherent detector, which combines all frames and beam dwells of a revisit, and a non-coherent detector, robust to inter-frame clock phase drift. The algorithms leverage per-dwell clutter whitening and the adaptive coherence estimator (ACE) detection statistic to improve performance. Both detectors estimate the range, radial velocity, and azimuth of targets in real time on a laptop graphics processing unit (GPU). Measurements with an FR1 ISAC proof-of-concept system based on a commercial base station show that the coherent detector outperforms the non-coherent one and recovers the full track.
\end{abstract}

\noindent

\IEEEpeerreviewmaketitle

\vspace{-0.2cm}
\section{Introduction}
\IEEEPARstart{A}{s} the \ac{6G} of mobile communications draws closer to its expected release date, practical validation of its expected main use cases becomes increasingly necessary. This helps to ensure that the real-world deployment is effective within the constraints of commercial telecommunications hardware. One of the key use cases for \ac{6G} is \ac{ISAC}, which consists of using the mobile network infrastructure to perform sensing tasks, such as the detection and estimation of the spatial parameters of passive {\chg targets \cite{itu2023m2160,liu2022isac}.} The widespread deployment of mobile \acp{BS}---initially intended exclusively for communications coverage---makes them the ideal option for providing widespread sensing capabilities at minimal {\chg cost \cite{wild2021jcas}.\par}

Within the applications of \ac{ISAC}, drone detection has been receiving substantial attention from the main mobile infrastructure providers, research institutions, and defense {\chg agencies \cite{song2025uavisac,dt_isac}.} Keeping in mind that drones are being extensively used for warfare, effectively solving the drone detection problem is fundamental for securing critical airspace and, consequently, {\chg infrastructure \cite{guvenc2018drones,brown2023cuas}. Common approaches for drone detection include dedicated radars, passive \ac{RF} sensing, and acoustic and optical sensors \cite{guvenc2018drones}, all of which require deploying dedicated sensing hardware.\par}

\ac{ISAC} drone detection inherently poses multiple challenges. First, sensing beam allocation {\chg requires} a well-designed scheme for harmonious coexistence of sensing and communications. For example, \ac{TDD} solutions lead to aliasing in the Doppler domain{\chg \cite{tdd_isac},} which needs to be addressed. Another relevant issue {\chg is the} operating frequency band. Operating in \ac{FR2} {\chg (24.25--71~GHz) has limited coverage per site \cite{rappaport2013mmwave}, whereas \ac{FR1} benefits from the existing macro-cell coverage. However, the limited bandwidth and wide beams in \ac{FR1} leave a small drone competing with strong ground clutter.\par}

In this work, we propose two algorithms that address the problem of monostatic \ac{ISAC} drone detection and parameter estimation. {\chg Both operate within the resource and beam allocation imposed by the communications system, using only the sensing symbols of a \ac{TDD} frame.} Our first and main approach leverages coherent combining, clutter whitening, and a modification of the \ac{ACE} \cite{ace}, which has \ac{CFAR} for properly whitened clutter. The other algorithm is a non-coherent estimator, robust to clock phase drift, which might be present in quasi-monostatic setups. The algorithms are conducive to implementation in a \ac{GPU}, which allows them to operate in {\chg real time.} We validate our methods with real-world measurement data from a quasi-monostatic \ac{FR1} \ac{PoC} setup and show that the coherent method consistently detects the drone and {\chg allows recovering its full track while simultaneously rejecting clutter.}

\section{System Model}\label{sec:model}
Consider a (quasi-)monostatic sensing scenario in which the \ac{BS} transmits $N_\text{C}\times N_\text{S}$-sized \ac{OFDM} frames with symbols {\chg $s_{ntf}$}; where $N_\text{C}$ and $N_\text{S}$ are the number of subcarriers and \ac{OFDM} symbols, respectively. Additionally, $n$, $t$, and $f$ denote the subcarrier, \ac{OFDM} symbol, and frame indices, respectively. The \ac{Tx} and \ac{Rx} arrays are digitally steered \acp{ULA} with identical steering. Assuming only single bounce reflections, then the response at the \ac{AoA}/\ac{AoD} $\psi$ is
\begin{gather}
    \alpha(\psi, \Bar{\psi}) =  \text{AF}^2(\sin \psi, \sin \Bar{\psi}) , \\
    \chg \text{AF}(\sin \psi, \sin \Bar{\psi}) = \sum^{N_\text{ant}-1}_{u=0} e^{-j 2\pi d (\sin \psi - \sin \Bar{\psi}) u},
\end{gather}
where {\chg $N_\text{ant}$ is the number of antenna elements,} $d$ is the antenna spacing in carrier wavelengths and $\Bar{\psi}$ is the steering direction. We call $\alpha$ and $\text{AF}$ the two-way and one-way array responses, respectively. \par

Let a \ac{CPI} consist of $F$ frames, which are divided into $B$ sets of beam \emph{dwells} $\mathcal{B}_g$ of equal size $F_g = |\mathcal{B}_g|$, such that $\sum^{B-1}_{g=0} F_g = F$. We say that a frame $f$ belongs to dwell $g$ if $f\in\mathcal{B}_g$. Thus, we define the mapping, $g(f) = \{ g \text{ where } f \in \mathcal{B}_g \}$\footnote{We consider that a frame can only belong to a single beam.}. A beam dwell is defined as the number of frames over which the \ac{Tx} and \ac{Rx} beamformers are held constant. The \ac{CPI} period containing $B$ distinct beam dwells is called a \emph{revisit}. Each frame is divided into $F_\text{sub}$ sub-frames of $N_\text{sf} = 70$ symbols, which are further subdivided into $N_\text{sens}$ sensing symbols and $N_\text{comm}$ communications symbols. This corresponds, for example, to a \ac{TDD} setup in which the \ac{BS} senses during $N_\text{sens}$ downlink symbols, while the remaining $N_\text{comm}$ symbols, e.g., uplink symbols, are not available for sensing. We assume that the beam is fixed and known only during the sensing symbols. Thus, effectively, for each frame we have an active set of sensing symbols, also called a \textit{mask}, defined as
\begin{multline}
    \mathcal{S} = \{ t = q N_\text{sf} + r : q = 0,\dots,F_\text{sub}-1 \\ \text{ and } r = 0,\dots,N_\text{sens}-1  \},
\end{multline}
with the total active symbol count being $N_\text{A} = |\mathcal{S}|$. The elements of $\mathcal{S}$ are indexed as $t^s_0<t^s_1<\dots<t^s_{N_\text{A}-1}$ and also organized in a vector $\mathbf{t}_\mathcal{S} = [t^s_0, \dots, t^s_{N_\text{A}-1}]$. \par

At frame $f$, the received signal is given by
\begin{equation}
    Y^\text{CSI}_{ntf} = H_{ntf} s_{ntf} + z^\text{AWGN}_{ntf},
\end{equation}
where $H_{ntf}$ are the elements of the clutter+target channel matrix $\textbf{H}_f$ and $z^\text{AWGN}_{ntf}\sim \mathcal{CN}(0,\sigma^2_z)$ is \ac{AWGN}. The channel matrix can be expressed as
\begin{equation}
    \chg H_{ntf} = \sum^L_{\ell=1} \gamma_\ell \alpha_{g(f)}(\psi_\ell) e^{j((t + t_f/T_\text{s})\omega_{2\ell} - n\omega_{1\ell})} + z_{ntf},
\end{equation}
where $\gamma_\ell$ is the complex path gain, $\alpha_{g(f)}(\psi_\ell) = \alpha(\psi_\ell, \psi_{g(f)})$ is the two-way array response in frame $f$,
$\psi_{g(f)}$ is the steering at dwell $g(f)${\chg , $t_f$ is the start time of frame $f$}, $\omega_{1\ell}$ and $\omega_{2\ell}$ are, respectively, the delay and Doppler angular frequencies, and $z_{ntf}$ is the clutter component. The former ones are given by
\begin{equation*}
    \omega_{1}(\rho) = \frac{4 \pi f_\text{sc} \rho}{c_0},\quad\quad \omega_{2}(\nu) = \frac{4 \pi f_\text{c} T_\text{s} \nu}{c_0},
\end{equation*}
where $f_\text{sc}$ is the subcarrier spacing, $\rho$ is the range, $c_0$ is the speed of light, $f_\text{c}$ is the carrier frequency, $T_\text{s}$ is the \ac{OFDM} symbol time, and $\nu$ is the radial {\chg velocity, positive for approaching targets.\par}

From the active symbols and subcarriers, the channel matrix is estimated by dividing the received signal by the active resource grid, i.e., $\Tilde{H}_{ntf} = Y^\text{CSI}_{ntf}/s_{ntf}$ for $t\in\mathcal{S}$ and $\Tilde{H}_{ntf} = 0$ otherwise. Then, a cropped periodogram $\mathbf{Y}_f \in \mathbb{C}^{N_\text{R} \times N_\text{D}}$ is computed for each frame by taking a windowed 2D-\ac{DFT} {\chg (inverse over subcarriers and forward over symbols)}~\cite{OFDM_radar} of the estimated \ac{CSI} matrix and cropping the $N_\text{R} \times N_\text{D}$ range$\times$Doppler map {\chg (centered around 0 in Doppler),} i.e.,
\begin{equation}
    \mathbf{Y}_f = \left[\text{2DFT}\left\{  \mathbf{w}_\text{R} \mathbf{w}_\text{D}^T \odot \Tilde{\mathbf{H}}_f \right\} \right]_{\mathcal{R} \times \mathcal{D}},
\end{equation}
where $\mathbf{w}_\text{R/D}$ denotes the window over range/Doppler (e.g., Chebyshev, Hamming, etc.) and $[\cdot]_{\mathcal{R} \times \mathcal{D}}$ denotes cropping to the relevant range and Doppler regions  $\mathcal{R}$ and $\mathcal{D}$, respectively. The periodogram stack $\{\mathbf{Y}_f\}_{f=0,\dots, F-1}$, as well as knowledge of the beams and system parameters, is the available data for target sensing.

{\chg \section{Detection and Parameter Estimation}\label{sec:estim}}
The proposed estimator fits a parametric clutter model per beam dwell assuming that the clutter covariance matrix admits a separable Kronecker structure over delay and Doppler. This fitted clutter model is  used to whiten the cropped periodograms. The estimation is subsequently performed entirely in the whitened domain. As the detection statistic, we use the \ac{ACE} \cite{ace}, which is scaling-invariant and provides \ac{CFAR} with respect to the clutter covariance\footnote{The CFAR property strictly holds only if the whitened clutter is Gaussian, which requires Gaussian clutter, a fitted covariance family that represents the real clutter covariance, and correctly modeled inter-frame clutter dependence.}.\par

Differently from earlier methods such as \ac{SAGE} \cite{sage,ch_est_pinto} and \mbox{RiMAX} \cite{rimax}, the proposed estimator does not perform successive cancellation. Instead, we perform a procedure which leverages regional maxima characterization with alias-aware \ac{NMS} to address the \ac{PSF} artifacts. This has the disadvantage of underdetecting closely spaced targets, since a weaker close target gets buried under a stronger peak. However, the approach was chosen for its computational efficiency, {\chg since} cancellation can be expensive when many targets are present.

\subsection{Clutter Fit}
We will assume that the clutter covariance respects a Kronecker structure over the range and Doppler frequency for each dwell $g$, i.e., vectorizing any frame from dwell $g$ by stacking constant Doppler columns yields
\begin{equation}
    \chg \mathbf{R}_g = \E{\text{vec}\{\mathbf{Y}_f\} \text{vec}\{\mathbf{Y}_f\}^H} = \mathbf{G}_{\text{D},g} \otimes \mathbf{G}_\text{R}\text{, for }f\in \mathcal{B}_g,
\end{equation}
where {\chg $\mathbf{G}_{\text{D},g}$} and $\mathbf{G}_\text{R}$ are, respectively, the Doppler and range clutter covariances.
We let the range covariance represent model based effects of \ac{DFT} windowing and periodogram cropping, while the Doppler covariance is computed from a parametric fit to data.

\subsubsection{Range Covariance}
The range covariance matrix is taken to be simply the squared range-window autocorrelation matrix. Thus, it is a Toeplitz covariance given by
\begin{equation}
    \chg [\mathbf{G}_\text{R}]_{mn} = \sum^{N_\text{C}-1}_{i=0} [\textbf{w}_{\text{R}}]^2_i e^{j 2\pi (i+n_0)(k_m - k_n)/N_\text{fr}} + \epsilon\delta_{mn},
\end{equation}
where {\chg $N_\text{fr}$ is the range \ac{DFT} size,} $n_0$ is a subcarrier offset, $k_{m}$ is the index of the cropped range bin $m$, and $\epsilon\delta_{mn}$ (where $\delta_{mn}$ is the Kronecker delta) is diagonal loading for conditioning.

\subsubsection{Doppler Sample Covariance}
We now compute a sample covariance for the clutter Doppler. Denote {\chg by $\mathbf{y}_{fi} \in \mathbb{C}^{N_\text{D}}$ the Doppler samples of range bin} $i$ of frame $f\in\mathcal{B}_g${\chg , i.e., $[\mathbf{y}_{fi}]_k = [\mathbf{Y}_f]_{ik}$, with $\E{\mathbf{y}_{fi} \mathbf{y}^H_{fi}} = [\mathbf{G}_\text{R}]_{ii} \mathbf{G}_{\text{D},g}$}. This means that we can estimate the Doppler sample covariance as
\begin{equation}
    \chg \widehat{\mathbf{G}}_{\text{D},g} = \frac{\sum_{f\in \mathcal{B}_g}\sum_{i\in\mathcal{K}_f}  \mathbf{y}_{fi} \mathbf{y}^H_{fi} }{ \sum_{f\in \mathcal{B}_g}\sum_{i\in\mathcal{K}_f} [\mathbf{G}_\text{R}]_{ii} },
\end{equation}
where $\mathcal{K}_f$ are the rows of a lower energy quantile (e.g., the lower 90th percentile), deliberately chosen to avoid targets biasing the estimates.

\subsubsection{Doppler Covariance Parametric Fit}

The slow-time (CSI domain Doppler) covariance is defined as the Billingsley \ac{ICM} model \cite{bllingsley}, omitting dependence on $g$
\begin{equation}
    \mathbf{R}(\boldsymbol{\theta}) = \sigma^2_0 \mathbf{I}_{N_\text{A}} + \phi_\text{DC} \boldsymbol{1}_{N_\text{A}} \boldsymbol{1}^H_{N_\text{A}} +
    \phi_\text{AC} \mathbf{T}_\text{AC}(t_h),
\end{equation}
where $\boldsymbol{\theta} = (\sigma^2_0, \phi_\text{DC}, \phi_\text{AC}, t_h)$ are the clutter parameters, and
$\mathbf{T}_\text{AC}(t_h)$ is the Lorentzian \ac{ICM} autocorrelation given by
\begin{equation}
    [\mathbf{T}_\text{AC}(t_h)]_{ab} = \frac{t^2_h}{t^2_h + (t^s_a - t^s_b)^2 },
\end{equation}
in which $t^s_a$ and $t^s_b$ refer to the time of the $a$th and $b$th active symbols, respectively. We now define the windowed \ac{DFT} matrix over the active symbol support $\mathbf{F}_\mathcal{S}\in\mathbb{C}^{N_\text{D} \times N_\text{A}}$ as
\begin{equation}
    [\mathbf{F}_\mathcal{S}]_{mn} = [\textbf{w}_{\text{D}}]_{t^s_n} e^{-j2\pi \frac{m t^s_n}{N_\text{fd}}},\text{ where }t^s_n\in\mathcal{S}
\end{equation}
where $N_\text{fd}$ is the Doppler-\ac{DFT} size. The slow-time covariance is then converted into an alias-aware Doppler covariance
\begin{equation}
    \begin{split}
    \chg \mathbf{G}_\text{D}(\boldsymbol{\theta}) &\chg = \mathbf{F}_\mathcal{S} \mathbf{R}(\boldsymbol{\theta}) \mathbf{F}^H_\mathcal{S} + \sigma^2_0 \mathbf{I}_{N_\text{D}} \\
    \chg &\chg = \sigma^2_0 (\mathbf{M}_0 + \mathbf{I}_{N_\text{D}}) + \phi_\text{DC} \mathbf{M}_\text{DC} + \phi_\text{AC} \mathbf{M}_\text{AC} (t_h),
    \end{split}
\end{equation}
where {\chg the added floor $\sigma^2_0 \mathbf{I}_{N_\text{D}}$ keeps $\mathbf{G}_\text{D}$ invertible, since \mbox{$\mathbf{M}_0 = \mathbf{F}_\mathcal{S}\mathbf{F}^H_\mathcal{S}$} is ill-conditioned, and} the mask-aware transformed covariance components are \mbox{$\mathbf{M}_\text{DC} = \mathbf{F}_\mathcal{S} \boldsymbol{1}_{N_\text{A}} \boldsymbol{1}^H_{N_\text{A}} \mathbf{F}^H_\mathcal{S}$} and \mbox{$\mathbf{M}_\text{AC}(t_h) = \mathbf{F}_\mathcal{S} \mathbf{T}_\text{AC}(t_h) \mathbf{F}^H_\mathcal{S}$}. The Doppler clutter covariance parameters $\boldsymbol{\theta}_g$ are fit per-dwell by first performing a joint \ac{LS} fit over $(\sigma^2_0, \phi_\text{DC})$ {\chg with the bases $(\mathbf{M}_0, \mathbf{M}_\text{DC})$}, followed by another \ac{LS} fit over $t_h$ over the residual from the previous fit\footnote{The parameter $\phi_\text{AC}$ has a closed form expression as a function of $t_h$.}.

\subsection{Whitening and Matched Filtering}
We perform two-sided whitening to the frames $\mathbf{Y}_f$ by applying the lower Cholesky factors to their corresponding directions
\begin{equation}
    \chg \Tilde{\mathbf{Y}}_f = \mathbf{L}^{-1}_\text{R} \mathbf{Y}_f \mathbf{L}^{-T}_\text{D,g}\text{, for }f\in\mathcal{B}_g,
\end{equation}
where {\chg $\mathbf{G}_\text{R} = \mathbf{L}_\text{R} \mathbf{L}^H_\text{R}$} and $\mathbf{G}_\text{D,g}(\boldsymbol{\theta}_g) = \mathbf{L}_\text{D,g} \mathbf{L}^H_\text{D,g}$. The whitening is also similarly applied for the candidate delay and Doppler atoms (matched filters). From the considered model, a candidate range atom matched to range $\rho$ is given by
\begin{equation}
    \chg \mathbf{a}_\text{R}(\rho) = \left[ N_\text{fr} \text{IDFT}\left( \mathbf{w}_\text{R} \odot e^{-j\omega_1(\rho) \mathbf{n}} \right) \right]_\text{R} \in \mathbb{C}^{N_\text{R}},
\end{equation}
where $\mathbf{n}$ is the $N_\text{C}$-sized vector of occupied subcarrier indices zero-padded to the range \ac{DFT} size $N_\text{fr}$. The candidate Doppler atom at speed $\nu$ is given by
\begin{equation}
    \mathbf{a}_\text{D}(\nu) = \mathbf{F}_\mathcal{S} e^{j\omega_2(\nu)\mathbf{t}_\mathcal{S}} \in \mathbb{C}^{N_\text{D}}.
\end{equation}
The range and Doppler atoms are used to construct the rank-1 atom given by $\mathbf{a}_\text{R}(\rho) \mathbf{a}^T_\text{D}(\nu) \in \mathbb{C}^{N_\text{R} \times N_\text{D}}$, which  for dwell $g$ is subsequently whitened, leading to
\begin{equation}
    \chg \Tilde{\mathbf{A}}_g(\rho,\nu) = \mathbf{L}^{-1}_\text{R} \mathbf{a}_\text{R}(\rho) \mathbf{a}^T_\text{D}(\nu) \mathbf{L}^{-T}_\text{D,g}.
\end{equation}
A matched filter operation is described by the Frobenius inner product {\chg $\langle \mathbf{A}, \mathbf{B} \rangle_F = \text{tr}(\mathbf{A}^H\mathbf{B})$} between a frame and its corresponding atom
\begin{equation}
    s_f(\rho,\nu) = \left\langle \Tilde{\mathbf{A}}_{g(f)}(\rho,\nu), \Tilde{\mathbf{Y}}_f \right\rangle_F.
\end{equation}

\subsection{Detection Statistics}
From the matched filter operation, we then define the per-frame \ac{ACE} as
\begin{equation} \label{eq:per_frame_ace}
    \Lambda_f(\rho,\nu) = \frac{|s_f(\rho,\nu)|^2}{ \| \Tilde{\mathbf{Y}}_f \|^2_F \| \Tilde{\mathbf{A}}_{g(f)}(\rho,\nu) \|^2_F },
\end{equation}
bounded within $[0,1]$ by the {\chg Cauchy--Schwarz} inequality. From (\ref{eq:per_frame_ace}), we define two detection statistics: a \ac{CO} estimator and a {\chg \ac{NC} estimator.} The \ac{NC} approach is robust to cross-frame \ac{CPE} and unknown frame timing. On the other hand, the \ac{CO} approach leverages phase coherence between frames (if it exists to a sufficient degree) to improve \ac{SCNR} and Doppler resolution. 
We focus on the coherent estimator, but we nonetheless present the noncoherent version and briefly explore it in the Section~\ref{sec:numres}.

\subsubsection{Coherent Estimator (CO)}
The contiguous frames belonging to a single dwell can be coherently combined by keeping track of the phase across frames. The frames of a dwell are coherently combined into matrices $\Bar{\mathbf{Y}}_g$. The dwells-collapsed frames are subsequently grouped in the tuple $\mathcal{Y} = \left(\Bar{\mathbf{Y}}_0, \dots, \Bar{\mathbf{Y}}_{B-1}\right)$. The tuples can be treated as usual tensors, thus we define the coherent \ac{ACE} using the standard tensor Frobenius inner-product and its induced norm
\begin{equation} \label{eq:co_ace}
    \Lambda_\text{CO}(\boldsymbol{\xi}) = \frac{ \left| \left\langle \mathcal{Y} , \mathcal{A}(\boldsymbol{\xi}) \right\rangle_F \right|^2  }{\left\|\mathcal{Y}  \right\|^2_F \left\| \mathcal{A}(\boldsymbol{\xi}) \right\|^2_F},
\end{equation}
where $\mathcal{A}(\boldsymbol{\xi})$ is the corresponding atom tuple and $\boldsymbol{\xi} = (\psi,\rho,\nu)$.

\subsubsection{Non-Coherent Estimator (NC)}
Similarly, we define the non-coherent estimator (or frame-coherent estimator) as the beam-pattern power weighted {\chg sum} of the per-frame \ac{ACE} statistics
\begin{gather} \label{eq:nc_ace}
    \chg \Lambda_\text{NC}(\boldsymbol{\xi}) = \frac{1}{C} \sum^{F-1}_{f=0} |\alpha_{g(f)}(\psi)|^2 \Lambda_f(\rho,\nu),\\
    \chg C = \max_{\psi} \sum^{F-1}_{f=0} |\alpha_{g(f)}(\psi)|^2,
\end{gather}
{\chg where the constant $C$ depends only on the beam geometry. With the per-frame statistics of (\ref{eq:per_frame_ace}), $\Lambda_\text{NC}$ is within $[0,1]$.}

\subsubsection{Localized Detection Statistics}
\begin{figure*}[!t]
\chg
\centering
\includegraphics[width=1\textwidth]{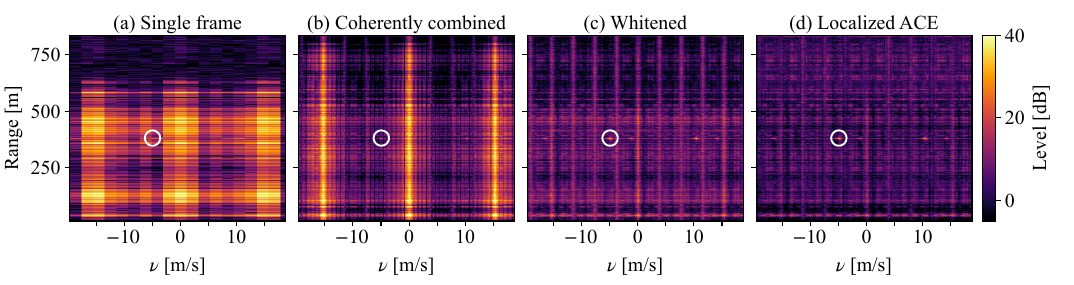}
\caption{Processing steps of the coherent estimator (CO) on one revisit with the drone (circle) at about 380~m and $-5$~m/s: (a) periodogram $|\mathbf{Y}_f|^2$ of a single frame of the dwell pointing closest to the drone; (b) coherent combination of all frames and dwells of the revisit without whitening; (c) the same with whitening; (d) localized \ac{ACE}. Panels (b)--(d): maximum over $\psi$, without the zero-Doppler guard band; dB relative to each panel's 20th percentile.}
\label{fig:progression}
\end{figure*}

The provided statistics take the whole periodogram data into account both in the numerator and in the denominator (normalization). This means that the total clutter power is taken into account for detection, not only the local clutter behavior around a particular path. Thus, if the whitener is not ideal, then unmodeled residual clutter will impact detection performance negatively. On the other hand, if only the local clutter is taken into account, then the detector becomes more robust to heterogeneous clutter distributions over $(\rho,\nu)$. For this reason, we define a local denominator \ac{ACE}. For the coherent {\chg case, for example,} we exchange $\|\mathcal{Y}\|^2_F$ in the denominator of (\ref{eq:co_ace}) for $E_\text{loc}$, such that
\begin{equation}
    E_\text{loc} = \frac{N_\text{R}N_\text{D}}{\left| \mathcal{T}(\rho,\nu) \right| } \!\!\!\!\!\!\!\!\!\!\!\!\!\!\!\!\!\!\! \sum_{\quad\quad\quad(k_1,k_2) \in \mathcal{T}(\rho,\nu)} \!\!\!\!\!\!\!\!\!\!\!\!\!\!\!\!\!\!\!\! \Bar{\mathcal{Y}}(k_1,k_2),
\end{equation}
where $\Bar{\mathcal{Y}}(k_1,k_2) = \sum^{B-1}_{g=0} | \left[ \Bar{\mathbf{Y}}_g  \right]_{k_1,k_2} |^2$, and $\mathcal{T}(\rho,\nu)$ is the set of range and Doppler indices centered at the bin that contains {\chg $(\rho,\nu)$} (excluding some guard bins in the center to minimize the influence of the target in the normalization). {\chg For the non-coherent estimator, the denominator $\| \Tilde{\mathbf{Y}}_f \|^2_F$ of each $\Lambda_f$ in (\ref{eq:nc_ace}) is replaced analogously by the local energy of frame $f$. Both detectors and their thresholds use these localized statistics, which are not bounded by one. Fig.~\ref{fig:progression} illustrates the processing steps of the coherent estimator on a measured revisit (see Section~\ref{sec:numres}): the drone, not visible under the clutter in a single frame, emerges after coherent combining and whitening.}

\subsection{Search and Admission} \label{sec:admission}

The respective \ac{ACE} field is densely sampled over $(\psi,\rho,\nu)$ to construct an \ac{ACE} tensor $\Lambda_{lmn} = \Lambda(\psi_l, \rho_m, \nu_n)$. The field is sampled according to the resolution of each dimension. Thus, the coherent estimator requires more samples, since it has an associated resolution gain of $F$ in $\nu$. For computational efficiency reasons, if some resolution gain can be forgone, the grid does not need to be constructed with $F N_\text{D}$ bins. The elements $(l,m,n)$ that satisfy a precomputed detection threshold $\Lambda_{lmn}>\eta$ and are outside a zero-Doppler guard {\chg band\footnote{Including the mask-induced zero-Doppler aliases.}, i.e., $|\nu_n|\geq\nu_\text{guard}$, are then kept} in a set $\mathcal{X}^\text{thld}$. Elements in $\mathcal{X}^\text{thld}$ are subjected to \ac{NMS} and postprocessing. The surviving elements are kept in the tuple $\mathcal{X}^\star$, where the main Doppler lobe is {\chg kept whenever it lies outside the guard band.} Under heavy blanking, the main lobe and its aliases effectively belong to an equivalence class, so the aliases can be recomputed from the minimum $\nu$ element.\par

The detected elements are then refined using an alternating coordinate descent {\chg procedure,} such as in \cite{sage,ch_est_pinto}. This procedure improves the resolution of the already detected points beyond the original grid. The coordinate descent is performed on {\chg $|\langle \mathcal{Y}, \mathcal{A}(\boldsymbol{\xi}) \rangle_F|^2/\|\mathcal{A}(\boldsymbol{\xi})\|^2_F$ for the coherent estimator and on the localized version of (\ref{eq:nc_ace}) for the non-coherent estimator. Refined detections that lie within a small range, Doppler, and angle tolerance of each other are then merged.}

\subsection{Algorithm}
A high level description of the algorithm is provided in Algorithm~\ref{alg:main_alg}. The {\chg two detectors differ mainly} in their detection statistic and {\chg in the Doppler density over which} $\Lambda_{lmn}$ is computed. Naturally, different detection statistics need different detection thresholds to satisfy a particular probability of false alarm.

\begin{algorithm}[!htbp]
\caption{{\chg Proposed one-shot detection algorithm}}
\label{alg:main_alg}
\begin{algorithmic}[1]
    \Procedure{Main}{$(\mathbf{Y}_f)_{f=0,\dots,F-1}$}
    \State Fit a clutter model for each dwell $g$;
    \State Whiten the frames $\Tilde{\mathbf{Y}}_f$;
    \State Compute the relevant \ac{ACE} cube $\Lambda_{lmn}$ as described in Section~\ref{sec:admission};
    \State Regional maxima detection and greedy descending value \ac{NMS}, yielding $\hat{L}$ points;
    \For{$\ell = 0,\,\dots,\,{\chg \hat{L}-1}$}
        \State Refine $\mathcal{X}^\star_\ell$ yielding $\boldsymbol{\xi}_\ell$
    \EndFor
    \State Collect the refined parameters $\boldsymbol{\xi} = (\boldsymbol{\xi}_\ell)_{\ell=0,\dots,\hat{L}-1}$;
    \State \Return $( \boldsymbol{\xi}, \hat{L} )$
\EndProcedure
\end{algorithmic}
\end{algorithm}

{\chg \section{Detection Threshold}\label{sec:threshold}}
\begin{figure*}[!b]
\chg
\centering
\includegraphics[width=0.69\textwidth]{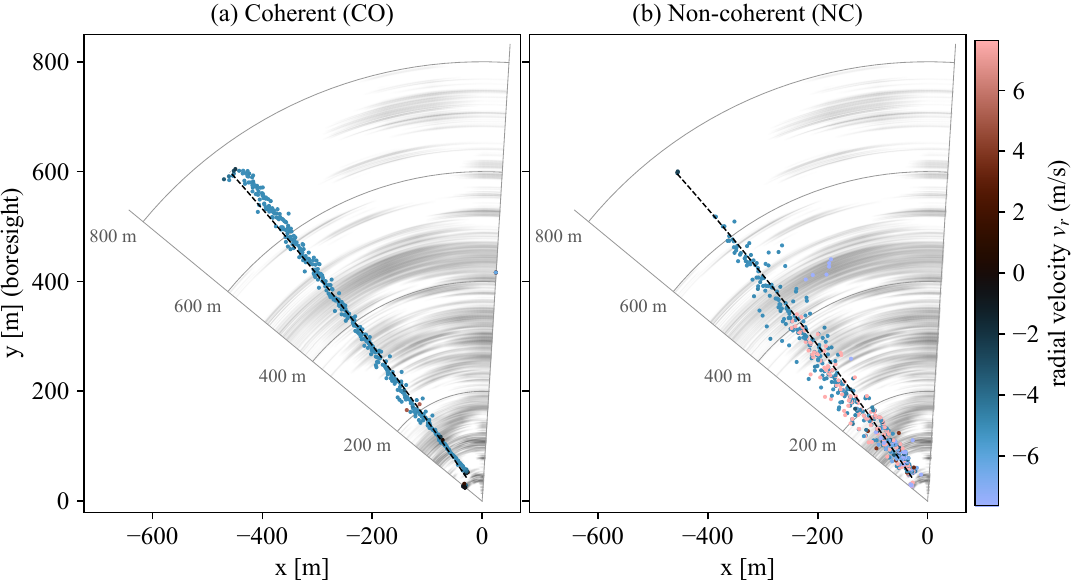}
\caption{Detections over the 413 revisits of the radial flight at the empirical threshold, top-down view: (a) coherent, (b) non-coherent detector. Color: estimated radial velocity (red: approaching, blue: receding; beyond the unambiguous $\pm 7.6$~m/s in the end colors). Dashed: fitted reference path. Gray: time-averaged clutter power, linearly interpolated in azimuth between the beam directions.}
\label{fig:wedge}
\end{figure*}
The $\eta$ threshold is computed and calibrated to a specific problem geometry. Since detections are computed from maxima over $\Lambda$, this means that the threshold must {\chg be} tuned to satisfy
\begin{equation}
    \mathbb{P}_{\mathcal{H}_0}  \left(\sup_{\boldsymbol{\xi}} \Lambda(\boldsymbol{\xi}) > \eta_\text{sup} \right) = \text{P}_\text{FA},
\end{equation}
which also necessarily leads to a higher threshold than satisfying $\mathbb{P}_{\mathcal{H}_0}  \left( \Lambda(\boldsymbol{\xi}) > \eta_\text{pt} \right) = \text{P}_\text{FA}$ at any particular point $\boldsymbol{\xi}$. We propose two approaches to calibrate $\eta$: from Monte Carlo theoretical $\mathcal{H}_0$ or from an empirical approach.

\subsection{Theoretical Threshold Calibration}
Draw $K$ {\chg target-free revisits from a reference Kronecker clutter model, then fit the clutter model and whiten each draw exactly as the measured data, so that $\Tilde{\mathbf{Y}}_f \sim \mathcal{CN}(\mathbf{0},\mathbf{I})$ holds up to the fitting error, which is thereby included in the threshold.} For each sample revisit, compute the {\chg localized} \ac{ACE} statistic and remove the zero-Doppler band as in Section~\ref{sec:admission}. For each draw $k$, compute the maximum of the admissible $\Lambda^\text{max}_k = \max_{\Omega} \Lambda^k_{lmn}$, where $\Omega$ is the admissible $(l,m,n)$ index set and $\Lambda^k_{lmn}$ is the corresponding \ac{ACE} statistic for draw $k$. Now, compute the mean $\mu_{\Lambda}$ and $\sigma_\Lambda$ of the vector $\boldsymbol{\Lambda}^\text{max} = ( \Lambda^\text{max}_k )_{k=0,\dots,K-1}$ and fit a Gumbel distribution using the method of moments
\begin{gather}
    \beta_\text{G} = \sigma_\Lambda \sqrt{6}/\pi \text{  (scale parameter)}, \\
    \mu_\text{G} = \mu_{\Lambda} - \gamma_\text{e} \beta_\text{G} \text{  (location parameter)},
\end{gather}
with $\gamma_\text{e}$ being the {\chg Euler--Mascheroni} constant. The resulting threshold is given by
\begin{equation}
    \eta(\text{P}_\text{FA}) =  \mu_\text{G} - \beta_\text{G}\log( -\log(1 - \text{P}_\text{FA}) ).
\end{equation}
Naturally, since actual clutter rarely fits the null hypothesis, $\text{P}_\text{FA}$ is just a reference parameter which does not correspond to the actual probability of false alarm.

\subsection{Empirical Threshold Calibration}
Take a reference capture with no targets and harvest all the candidate peaks per revisit. Then, choose $\eta$ such that a particular false alarm budget per revisit is satisfied. This leads to a more strict adherence to a true $\text{P}_\text{FA}$ value, with the downside of requiring training data for $\mathcal{H}_0$.

\section{Experimental Results}\label{sec:numres}

{\chg We evaluate both detectors on a measurement from an \ac{FR1} \ac{PoC} \ac{BS} operating at a carrier frequency of 3.93~GHz with 30~kHz subcarrier spacing \cite{3gpp.38.211} and $N_\text{C}=3276$ subcarriers (98.3~MHz, i.e., 1.53~m range resolution). Sensing uses $N_\text{sens}=6$ of every $N_\text{sf}=70$ symbols, i.e., a sensing overhead of 8.6\%, without windowing. The \ac{BS} steers an 8-element half-wavelength \ac{ULA} over $B=7$ beam dwells of $F_g=5$ frames ($F=35$), and one revisit lasts 0.35~s. A drone remains at about 60~m from the \ac{BS}, then flies approximately radially outbound at about 5~m/s, at an azimuth of about $-35^\circ$, to a range of about 750~m, and stops there. We evaluate the 413 revisits (141~s) of the radial flight.\par}

{\chg No independent ground truth---e.g., \ac{GNSS}---is available for this flight. Therefore, we fit a straight reference path, i.e., range and azimuth linear in time, to the receding detections of both detectors near the drone bearing. The fit is performed by robust M-estimation~\cite{huber1964robust} with a pseudo-Huber loss \cite{charbonnier1997edge} and is iteratively refitted on the gated detections. A detection is attributed to the drone if it lies within $\pm 25$~m in range and $\pm 15^\circ$ in azimuth of the path, which we call the \textit{drone gate}. All other detections with $|\nu| > 0.5$~m/s are called \textit{non-drone} detections. The detection rate is the fraction of revisits with at least one gated detection, and the scatter is the \ac{RMSE} of the gated detections about the path.\par}

{\chg For each detector, we use the theoretical threshold at $\text{P}_\text{FA}=10^{-3}$ ($K=150$ null revisits) and an empirical threshold set for 0.1 non-drone detections per revisit. Since no target-free capture of this scene is available, the empirical threshold is computed on the same capture. This means that, at the empirical threshold, both detectors are compared at an equal non-drone detection load.}

{\chg Fig.~\ref{fig:wedge} and Table~\ref{tab:results} show the results at the empirical threshold. It can be seen that the coherent detector detects the drone in nearly every revisit up to the end of the flight at about 750~m from the \ac{BS}, whereas the non-coherent detector detects it in about three quarters of the revisits and loses it beyond roughly 600~m, where it detects the drone in 9 of the 91 revisits against all 91 for the coherent detector. This happens because the coherent detector combines all frames and beam dwells of a revisit in phase, whereas the non-coherent detector only sums their per-frame statistics, which yields a smaller integration gain.\par}

{\chg At the theoretical threshold, the coherent and non-coherent detectors produce 0.002 and 0.09 non-drone detections per revisit, respectively. As discussed in Section~\ref{sec:threshold}, $\text{P}_\text{FA}$ is only a reference parameter, since the measured clutter does not follow the null model exactly. Nevertheless, the coherent detector deviates much less from it. \chg The range scatter about the reference path is below the 1.53~m range resolution for both detectors.\par}

{\chg Regarding computational cost, on an NVIDIA RTX 500 Ada Generation laptop \ac{GPU} (10 runs of all 456 revisits, excluding the first revisit of each run), the coherent detector takes 0.14~s per revisit on average (median 0.13~s, maximum 0.30~s) and the non-coherent detector 0.08~s (median 0.07~s, maximum 0.20~s). Thus, both detectors process every revisit within the 0.35~s revisit period.}

\begin{table}[!t]
\chg
\renewcommand{\arraystretch}{1.2}
\caption{Detection Performance on the FR1 Drone Measurement (413 Revisits of Radial Flight; Empirical Thresholds Set for 0.1 Non-Drone Detections per Revisit)}
\label{tab:results}
\centering
\footnotesize
\setlength{\tabcolsep}{3pt}
\begin{tabular}{llcc}
\hline
Threshold & Metric & CO & NC \\
\hline
Theoretical & Det. rate (\%) & 80 & 71 \\
($\text{P}_\text{FA}=10^{-3}$) & Non-drone/rev. & 0.002 & 0.09 \\
 & $\rho$ scatter (m) & 0.71 & 1.12 \\
\hline
Empirical & Det. rate (\%) & 99 & 73 \\
(0.1 non-drone/rev.) & $\rho$ scatter (m) & 0.65 & 1.10 \\
\hline
\end{tabular}
\end{table}

\section{Conclusion}\label{sec:conclusion}
{\chg We have proposed two clutter-whitened \ac{ACE} detectors for quasi-monostatic drone detection in \ac{FR1} \ac{ISAC}: a coherent detector, which combines all frames and beam dwells of a revisit, and a non-coherent detector, which is robust to inter-frame phase drift. Measurements with an \ac{FR1} proof-of-concept base station have shown that, at an equal non-drone detection load, coherent combining detects the drone more consistently, and that both detectors run in real time on a laptop \ac{GPU} with low sensing overhead and without dedicated radar hardware. Future work includes evaluation against \ac{GNSS} ground truth on multiple flights and addressing inter-frame phase drift.}

\section*{Acknowledgements}
The authors would like to thank Artjom Grudnitsky and Tomasz Sztuka for setting up the PoC and providing the measurement data.
The work has been partially funded by the Federal Ministry of Research, Technology and Space of Germany in the project “SENSATION” under grant number 16KIS2523K. The work was also supported by the Research Council of Finland 6G Flagship Program (Grant Number: 369116) and S6GRAN project (370559).

\ifCLASSOPTIONcaptionsoff
  \newpage
\fi

\bibliographystyle{IEEEtran}
\bibliography{IEEEabrv,biblio}

\end{document}